\documentclass[aip,jcp,graphicx,reprint]{revtex4-1}
\usepackage{graphicx} 
\usepackage{amsmath}

\newcommand\hfdim{HF$\ldots$HF~}
\newcommand\hfar{HF$\ldots$Ar~}
\newcommand\nndim{N$_2\ldots$N$_2$~}
\newcommand\codim{CO$\ldots$CO~}
\newcommand\waterdim{H$_2$O$\ldots$H$_2$O~}
\newcommand\methanear{CH$_4\ldots$Ar~}
\newcommand\watercoo{H$_2$O$\ldots$CO$_2$~}
\newcommand\coodim{CO$_2\ldots$CO$_2$~}
\newcommand\methanedim{CH$_4\ldots$CH$_4$~}
\newcommand\watermethane{H$_2$O$\ldots$CH$_4$~}

\newcommand\meohwater{CH$_3$OH$\ldots$H$_2$O~}

\draft 

\begin{document}

\title{The V30 Benchmark Set for Anharmonic Vibrational Frequencies of Molecular Dimers} 

\author{Johannes Hoja}
\email[]{johannes.hoja@uni-graz.at}

\author{A. Daniel Boese}
\affiliation{Department of Chemistry, University of Graz, Heinrichstraße 28/IV, 8010 Graz, Austria.}

\date{11 September 2024}

\begin{abstract}
Intermolecular vibrations are extremely challenging to describe but are the most crucial part for determining entropy and hence free energies, and enable for instance the distinction between different crystal-packing arrangements of the same molecule via THz spectroscopy. Herein, we introduce a benchmark data set --- V30 --- containing 30 small molecular dimers with intermolecular interactions ranging from exclusively van-der-Waals dispersion to systems with hydrogen bonds. All calculations are performed with the gold standard of Quantum Chemistry CCSD(T). We discuss vibrational frequencies obtained via different models starting with the harmonic approximation over independent Morse oscillators up to second-order vibrational perturbation theory (VPT2), which allows a proper anharmonic treatment including coupling of vibrational modes. 
However, large amplitude motions present in many low-frequency intermolecular modes are problematic for VPT2. In analogy to the often used treatment for internal rotations, we replace such problematic modes by a simple one-dimensional hindered rotor model. We compare selected dimers with available experimental data or high-level calculations of potential energy surfaces and show that VPT2 in combination with hindered rotors can yield a very good description of fundamental frequencies for the discussed subset of dimers involving small and semi-rigid molecules.
\end{abstract}


\maketitle

\section{Introduction}

The future of Computational Chemistry undoubtedly lies in the accurate description of intermolecular interactions: many of the modern fields of science, such as Nanotechnology or Biotechnology, are governed by intermolecular interactions, and their accurate description is one of the main challenges for the subject of Theoretical and Quantum Chemistry.

There was a lot of improvement in recent years concerning the calculation of the interaction energies.
First, advances in density functional theory\cite{Mardirossian2016,Verma2020,Furness2020} (DFT) 
and especially their dispersion descriptions\cite{Grimme2016,Hermann2017,Price2021,Vydrov2010}
 have pushed the limit of use further towards materials and liquids. Second, the applicability of the so-called gold standard in Quantum Chemistry, Coupled Cluster Theory including single, double, and perturbative triple excitations --- CCSD(T) --- has been extended to much larger molecules with the advent of localized methods, such as localized natural orbital Coupled Cluster LNO-CCSD(T)\cite{Nagy2019}
 or (domain-based) local pair natural orbital Coupled Cluster DLPNO-CCSD(T)\cite{Ma2018,Guo2018,Schmitz2016}.
 Hence, accurate energetics for non-covalent interactions can be obtained for increasingly large  and more complex systems utilizing the above-mentioned approximations.

The accurate computation of vibrational properties like infrared/Raman spectra or free energies, however, is a different story. 
As the studied molecular complexes become larger and larger, up to periodic systems, entropy and zero-point effects will start playing a larger role\cite{Grimme2012} --- even for structures\cite{Hoja2016,Dolgonos2019}.
Whereas zero-point effects can somewhat more easily be  approximated by using independent harmonic oscillators (harmonic approximation),
the accurate description of the entropy and hence free energy is definitely a challenge, since it mainly depends on the low-frequency vibrations.  In molecular complexes or molecular crystals, such modes involve intermolecular motions and capturing the associated subtle energy changes would require a more sophisticated treatment than the harmonic approximation.

There are a variety of approaches to improve upon the harmonic approximation and account also for anharmonic effects.
The simplest --- and probably most often used --- approach is to utilize the known general trend of the given method and simply scale all harmonic frequency by empirical factors\cite{Kesharwani2014,Merrick2007,Grev1991,Alecu2010}. While this may statistically somewhat improve the overall description, it does not provide any fundamental methodological improvement. 

A slightly more complicated approach is the replacement of the independent harmonic oscillators with independent Morse oscillators\cite{Morse1929,Dahl1988}. This can be done by fitting a Morse potential to energies of several structures, which have been displaced along normal-mode coordinates. While this approach can capture the individual anharmonic nature of some modes, there is no coupling between vibrational modes, which may be important. 

In contrast, the framework of molecular dynamics would in principle allow a straightforward evaluation of anharmonic vibrations. However, complications arise from the fact that molecular dynamic simulations may need very long simulation times to sufficiently scan the potential energy surface\cite{Ding2011}, especially for low-frequency vibrations, and also quantum nuclear effects may need to be included as well\cite{Rossi2016}.
Thus, approximations to simulate different vibrational modes with different displacements have been proposed\cite{Galimberti2021},
which nevertheless require long simulation times usually out of reach for post-Hartree-Fock methods.

One of the computationally less expensive ways to properly include anharmonic effects is second-order vibrational perturbation theory (VPT2)\cite{Mills1972,Csszr2011,Barone2005,Barone2010,Bloino2012,Barone2014,Bloino2016,Franke2021}.
Basically, the harmonic approximation is expanded by fully computing the third-order force constant matrix together with the diagonal part of the fourth-order force constant matrix, whereas the latter two are treated as perturbation of the harmonic system. This approach requires the calculation of Hessians for a number of geometries displaced along normal-mode coordinates. Within this methodology coupling between vibrational modes and also Coriolis couplings are included. However, occurrences of Fermi\cite{Fermi1931} or Darling Dennison resonances\cite{Darling1940} can cause numeric problems in the perturbation theory. Hence, several methods for resonance treatments have been proposed including DVPT2, VPT2+K\cite{Rosnik2013}, and GVPT2\cite{Piccardo2015}. In general, VPT2 approaches have successfully been used for isolated semi-rigid molecules, often relying completely or partially on DFT calculations\cite{DanielBoese2005,Boese2003,Boese2005}.
However, large-amplitude motions can often not correctly be described in the typically used cartesian normal coordinates\cite{Bloino2016}. Nevertheless, VPT2 was also successfully applied to some hydrogen-bonded dimers using CCSD(T) and larger hydrogen-bonded complexes using DFT\cite{Howard2014,Perlt2019,Buczek2021}.

Furthermore, there is also the vibrational self-consistent field (VSCF)\cite{Chaban1999,Monserrat2013,Kapil2019,Bowman1978} approach,
for which parts of the potential energy surfaces have to be calculated in a given coordinate system. Then, wave functions of the vibrational modes are solved self-consistently. This method is also quite dependent on the used coordinate system. In principle, proper internal coordinates would be needed for the correct description of intermolecular motions, but these are not easily computed and transformed, especially for molecular complexes. Such approaches can be further improved upon by vibrational configuration interaction (VCI) or vibrational coupled cluster (VCC) techniques\cite{Christiansen2004,Knig2015,Mathea2021,Knig2020}.

Finally, full potential energy surfaces can be calculated and then fitted by either analytic functions\cite{Braams2009,Qu2021,Carter2011} or also approximated by machine learning\cite{Dral2020,Lu2022}. This is, however, the most expensive --- but most accurate --- approach, as a huge number of points on the potential energy surface needs to be computed. Therefore, this approach quickly becomes intractable for larger systems due to the extremely unfavorable scaling with system size.

Herein, we want to benchmark and compare several vibrational approaches for molecular dimers. Therefore, we introduce a set containing 30 dimers of small molecules denoted as V30. This set should capture the whole range of intermolecular interactions and hence is not limited to hydrogen-bonded systems but also include dimers solely bound by van-der-Waals dispersion interactions. While vibrations of molecular dimers are certainly important for studying  for instance solvation effects, they could also provide vital information for the modeling of THz spectra of molecular crystals, which can be used to distinguish between different polymorphs in a non-destructive way. Therefore, THz spectra are vital for the pharmaceutical industry or for the detection of explosives and drugs in security screenings. Accurate vibrational properties for dimers could potentially directly be incorporated into a periodic description of molecular crystals by the utilization of a multimer embedding scheme\cite{Loboda2018,Hoja2023}. Such an approach could potentially also improve the accuracy of free energies needed for molecular crystal structure predictions\cite{Hoja2019}.

In addition to the ubiquitous harmonic approximation and simple Morse oscillators, we focus here on VPT2 approaches  because of their relative simplicity compared to other proper anharmonic treatments. In order to limit the error of the computed electronic energies and solely focus on the different vibrational approaches, we utilize the canonical gold standard CCSD(T) throughout this paper, paving the way for comparing more approximate methods to these very accurate frequencies later on. Especially for dimers without hydrogen bonds, several low-frequency modes involving large-amplitude motions cannot properly be described by canonical VPT2 methods when using displacements based on cartesian normal coordinates. For internal rotations, such modes are typically deperturbed or decoupled and then described by a hindered rotor model\cite{Bloino2012,Ayala1998,Dzib2021}. In analogy, we further test here the usefulness of augmenting VPT2 with a simple one-dimensional hindered rotor model for problematic intermolecular modes. For selected systems we then compare our obtained results with experimental data and available high-level calculations of potential-energy surfaces.

\section{Computational Methods}

\subsection{Systems and Geometries}

\begin{figure*}[ht]
\includegraphics[width=0.91\textwidth]{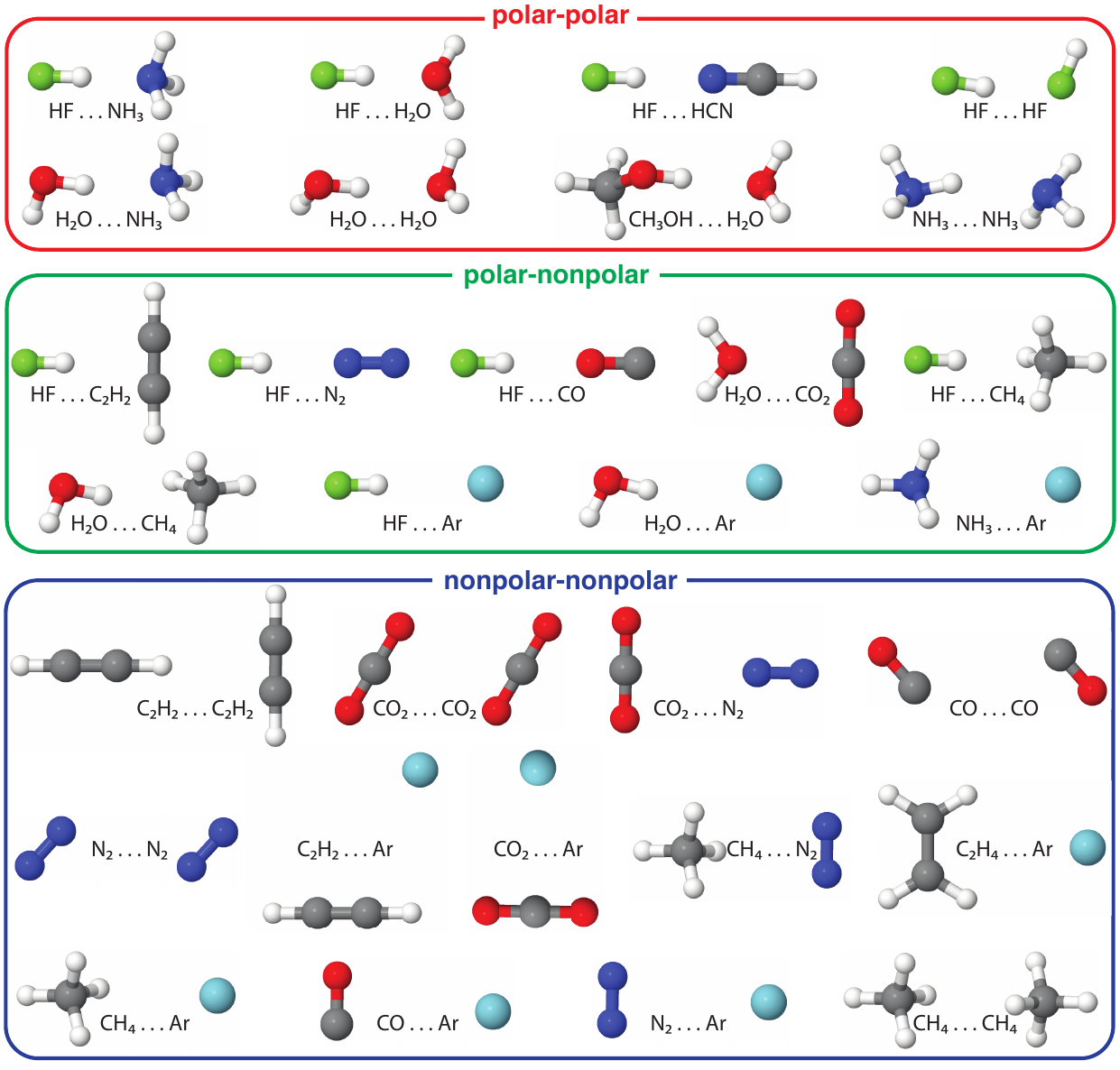}
\caption{\label{fig:set}Visualization of all molecules within the V30 benchmark set.}
\end{figure*}

In order to evaluate the performance of several approaches for calculating vibrations within molecular complexes, we have created a diverse benchmark set consisting of 30 small molecular dimers, which we label the V30 set (see Fig. \ref{fig:set}).
The system size was chosen so that anharmonic CCSD(T) calculations could still be performed.
The dimers within this set are either held together by van-der-Waals (vdW) dispersion interactions alone or also by hydrogen-bonding interactions. 
Therefore, we have divided V30 in three subsets: the first part contains dimers of two polar molecules, which geometrically form hydrogen bonds; the second part contains dimers formed by one polar and one nonpolar molecule, while the third part contains dimers of two nonpolar molecules.
Although carbon monoxide (CO) is formally a polar molecule, we classify it here as nonpolar due to its very weak dipole moment of only 0.1 Debye.

In order to minimize the error of the electronic energy in these benchmark calculations, all geometries and vibrational frequencies were calculated at the CCSD(T) level.
Throughout this paper, we are utilizing so-called \emph{heavy-augmented} correlation-consistent basis sets, \emph{i.e.}, cc-pVXZ\cite{Dunning1989} for hydrogen atoms and aug-cc-pVXZ\cite{Dunning1989,Kendall1992,Woon1993} for all non-hydrogen (heavy) atoms with X being T and Q. 
Howard \emph{et al.}\cite{Howard2014} have shown that these heavy-augmented basis sets perform very well for frequencies of the water dimer and the HF dimer. 
Herein, we abbreviate these basis sets with \emph{haXZ}. All dimers shown in Fig. \ref{fig:set} were optimized with CCSD(T)/haQZ utilizing Molpro  2020.2\cite{Molpro,Werner2011,Werner2020,Hampel1992,Deegan1994}. 
The frozen-core approximation was utilized throughout and energies were converged to $10^{-9}$ Hartree or better.
During geometry optimizations the maximal component of the gradient was generally converged to $10^{-6}$ Hartree/Bohr or better, only for N$_2\ldots$N$_2$ the gradient could only be converged to $5\times10^{-5}$ Hartree/Bohr. 

The optimized haQZ dimer structures are visualized in Fig. \ref{fig:set} and the coordinates are available in the Supporting Data\cite{SI}. 
In addition, the symmetry point groups of all structures are listed in Table S1 in the Supporting Information.
In the global minimum structure of the methanol-water complex \meohwater the water molecule acts as the hydrogen bond donor\cite{Moskowitz2001}. 
Herein, we have included a local minimum structure, in which the methanol molecule acts as hydrogen bond donor due to the favorable $C_S$ symmetry of that structure. 
The \watermethane complex has two stationary points in $C_S$ symmetry\cite{Qu2015}. For this system, we utilize the global minimum, in which one hydrogen atom from the water molecule is pointing towards the methane molecule (see Fig. \ref{fig:set}). 
In the secondary minimum (not considered here), one hydrogen atom from the methane is pointing towards in between of the water lone pairs.
For \nndim the potential energy surface is rather flat, leading to several possible local minima\cite{Jafari2005,Aquilanti2002}.
Here, we utilize a slanted parallel orientation, which is typically labeled as Z structure and represents the observed global minimum in Ref. \citenum{Jafari2005}. For \codim the global minimum is also a Z structure\cite{Dawes2013} with an anti-parallel alignment of the CO molecules. 

In order to further classify the discussed systems, all dimers within each of the three groups are ordered according to their magnitude of the sum of electrostatic and induction interactions compared to van-der-Waals dispersion interactions. 
This ratio is obtained from symmetry-adapted perturbation theory\cite{Jeziorski1994} by using DF-DFT-SAPT\cite{Hesselmann2005} calculations within Molpro on top of the CCSD(T)/haQZ geometries, which were performed with an asymptotic corrected PBE0 functional utilizing the aug-cc-pVQZ basis set.
The needed HOMO energies and (vertical) ionization potentials were calculated with PBE0/aug-cc-pVQZ.
In DFT-SAPT the interaction energy $E_{\rm SAPT}$ is decomposed into 
\begin{equation}
  \begin{split}
     E_{\rm SAPT} =  E_{\rm elst} +  E_{\rm exch} +  E_{\rm ind} +  E_{\rm exch-ind} + \delta_{\rm HF}\\  +  E_{\rm disp} +  E_{\rm exch-disp},   
  \end{split}
\end{equation}
where $E_{\rm elst}$ is the electrostatic energy, $E_{\rm exch}$ the first-order exchange energy, $E_{\rm ind}$ the second-order induction energy, $E_{\rm exch-ind}$ the second-order exchange-induction energy, $\delta_{\rm HF}$ an estimate for higher-order induction contributions obtained from Hartree-Fock, $E_{\rm disp} $ the second-order dispersion energy, and $E_{\rm exch-disp}$ the second-order exchange-dispersion energy. 
Fig. \ref{fig:set} shows for each category the systems line by line ordered by decreasing values of the term
\begin{equation}
    R^{\rm elst,ind}_{\rm disp} = \frac{E_{\rm elst} + E_{\rm ind} +  E_{\rm exch-ind} + \delta_{\rm HF}}{E_{\rm disp} +  E_{\rm exch-disp}}.
\end{equation}
The numeric values of this ratio and the components are available in the Supporting Information in Table S1.

\subsection{Harmonic and VPT2 Frequencies}

On top of the optimized dimers, harmonic frequencies were calculated with CCSD(T)/haQZ using CFOUR 2.1\cite{cfour,Matthews2020}, making use of the available analytic second derivatives. The convergence of the Hartee-Fock equations, the coupled cluster amplitudes, and linear equations was generally set to $10^{-10}$, but sometimes had to be reduced to $10^{-9}$ or $10^{-8}$ for systems containing C$_2$H$_2$ and C$_2$H$_4$, as well as for CO$_2\ldots$N$_2$. The largest calculated gradient in CFOUR for the structures optimized using Molpro was $2\times10^{-6}$ Hartree/Bohr. 

Next, anharmonic VPT2 calculations were performed in the same fashion utilizing CFOUR. The needed cubic and semi-diagonal quartic force constants can be obtained from Hessians of displaced structures\cite{DanielBoese2005,Boese2005,Barone2005}. Unless specified otherwise, we use herein the default displacement of $0.05$ in reduced normal coordinates\cite{Yagi2004} $q$, which are related to the cartesian normal coordinates $Q$ via
\begin{equation}
    q = \sqrt{\frac{\omega}{\hbar}}Q,
\end{equation}
with $\omega$ being the angular vibrational frequency.

The standard VPT2 frequencies and intensities (without any resonance treatment) were calculated up to two quanta (fundamentals, first overtones, and combination bands) using the \textsc{xguinea} module provided in CFOUR. In a next step, the resonances identified by the xcubic module of CFOUR were taken into account by diagonalizing a respective vibrational Hamiltonian\cite{Rosnik2013} using xguinea. This approach is then labeled VPT2+K. Furthermore, the calculated force constants were then imported in Gaussian 16\cite{g16} in order to calculate frequencies with the generalized vibrational perturbation theory (GVPT2)\cite{Barone2005}, which is an automated variational-corrected VPT2 method. 

For the largest systems --- \coodim, C$_2$H$_2\ldots$C$_2$H$_2$, \methanedim, CH$_3$OH$\ldots$H$_2$O --- only harmonic calculations could be performed with the haQZ basis set and therefore we utilized the cubic and quartic force constants from haTZ calculations. All obtained force constants and calculated frequencies are available in the Supporting Data\cite{SI} (Zenodo dataset) in a database format compatible with the Atomic Simulation Environment (ASE)\cite{ase-paper}. Infrared intensities are available for harmonic calculations, VPT2, and VPT2+K. Tables including all calculated fundamental vibrations are additionally also available in the Supporting Information. 

Initially, all calculations were performed by using the default atomic masses within CFOUR, which are those of the respective most-stable isotope. For better compatibility with ASE, we have adapted the quadratic and cubic force constants to the standard atomic weights\cite{Meija2016} (ASE default) utilizing CFOUR. Given the only very small differences in the resulting frequencies, we will only discuss in the main text results obtained with the standard atomic weights. However, force constants and obtained frequencies up to this point utilizing the masses of the most-common isotopes are provided in a database within the Supporting Data\cite{SI}.

\subsection{Hindered Rotors}

One rather big obstacle for accurate anharmonic frequencies for dimers (but also for flexible monomers)\cite{Bloino2016} is given by so-called large-amplitude motions. In such modes, the two molecules perform often a kind of hindered rotation against each other. In several such cases, VPT2 does not provide reliable results, since such motion cannot be considered a small perturbation any more. This can then lead to a large overestimation or underestimation of the vibrational frequency, or even lead to imaginary frequencies. 

Typically, such problematic modes are excluded from the VPT2 treatment and replaced by a one-dimensional model\cite{Bloino2016}. Such modes are then independent and do not have any couplings to other modes anymore. Herein, we attempt to deal with such problematic intermolecular large-amplitude motions in quite a simple way. In case the mode represents a rather pure hindered rotation around the respective monomer center of mass, we decouple this mode from all other modes and treat it as a one-dimensional hindered rotor as described below. In case the harmonic mode shows a significant movement of the monomer center of mass, we also decouple this mode, but simply use the harmonic value. This approach is then denoted by adding +R, leading to either VPT2+R, VPT2+K+R, or GVPT2+R. 

The Hamiltonian of a one-dimensional hindered rotor\cite{McClurg1997,Pfaendtner2007} is given by
\begin{equation}
    \hat{H}_{\rm HR} = -\frac{\hbar^2}{2I_{\rm red}}\frac{d^2}{d\theta^2} + V_{\rm HR}(\theta),
\end{equation}
where $I_{\rm red}$ is the reduced moment of inertia corresponding to the rotation axis and the potential\cite{Pfaendtner2007} $V_{\rm HR}(\theta)$ is given by
\begin{equation}
   V_{\rm HR}(\theta) =   \sum_{k} \left[a_k(1-\cos k\theta) + b_k \sin k\theta \right].
\end{equation}
Herein, we obtain this potential by rigid rotations of one monomer in steps of 10 degrees and a subsequent CCSD(T)/haQZ single point calculation. Then, the parameters of the potential are obtained via a least-squares fit. In order to avoid overfitting, the lowest order ($k$) capable of describing the correct potential shape is used.  In most cases the rotations were generated by rotating about the axis connecting the two centers of mass or about a corresponding perpendicular axis. In case both monomers are rotating about the same axis, both individual reduced moments of inertia $I_{\rm red}^A$ and $I_{\rm red}^B$ are combined\cite{Pfaendtner2007,Dzib2021} according to
\begin{equation}
    \frac{1}{I_{\rm red}} =  \frac{1}{I_{\rm red}^A} +  \frac{1}{I_{\rm red}^B}, 
\end{equation}
and we utilize the potential with the smaller rotation barrier. In case the motion of one monomer is dominant, we only consider this rigid rotation.

Next, we obtain the frequency corresponding to this hindered rotation\cite{McClurg1997} via
\begin{equation}
 \nu_{\rm HR} = \frac{1}{2\pi}\sqrt{\frac{k_{\rm HR}}{I_{\rm red}}}  ,  
\end{equation}
where $k_{\rm HR}$ corresponds to the curvature of the potential around one minimum
\begin{equation}
    k_{\rm HR} = \left(\frac{d^2V_{\rm HR}}{d\theta^2}\right)_{\theta = 0}.
\end{equation}

This curvature approach is best suited for larger rotational barriers between minima. In case we observe an energy barrier between distinct minima of less than 2 kJ/mol (or 170 cm$^{-1}$), we use the obtained potential to numerically solve the one-dimensional Schrödinger equation (see Hamiltonian above) and obtain the frequency from the energy differences between eigenvalues. A list of which mode was treated with which rotor model ("c" for curvature or "h" for Hamiltonian) is available for every system in the respective table in the Supporting Information (Tables S4 to S33) and also within the Supporting Data\cite{SI}. The latter also includes the utilized parameters for the respective $V_{\rm HR}(\theta)$ potentials.

\subsection{Morse Oscillators}

In addition to VPT2, we also evaluate the simplest anharmonic approach, which is given by independent one-dimensional Morse oscillators\cite{Morse1929,Dahl1988}. The Morse potential is given by
\begin{equation}
    V_{\rm M}(x) = D \left[1-e^{-a(x-x_0)}\right]^2,
\end{equation}
with
\begin{equation}
    \omega_0 = \sqrt{\frac{2a^2D}{\mu}}.
\end{equation}
 The parameters $a$, $D$, and $x_0$ describe the potential width, the well depth, and the minimum, respectively. Since an analytical solution exists for the Morse potential, the energy corresponding to the vibrational quantum number $v$ can be calculated according to
\begin{equation}
    E(v) = \omega_0 \left(v + \frac{1}{2}\right) - \omega_0\chi \left(v+ \frac{1}{2}\right)^2, 
\end{equation}
whith $\chi$ being the so-called anharmonicity constant
\begin{equation}
    \chi = \frac{\omega_0}{4 D}.
\end{equation}
In order to obtain the parameters for the Morse potential, every mode was displaced by the same amplitude as used for VPT2 scaled by $\pm 1$ and $\pm 0.5$, followed by a CCSD(T)/haQZ single point calculation on top of the so displaced structures. The frequencies for the fundamental vibration and the first overtone were then obtained via $\omega_0 - 2\chi \omega_0$ and $2\omega_0-6\chi \omega_0$, respectively. In this model, there is no coupling between modes, so the combination frequencies just list the sum of two fundamental modes. We note that Morse and hindered rotor fits were only performed with CCSD(T)/haQZ using standard atomic weights\cite{Meija2016}.

\begin{table*}
\setlength{\tabcolsep}{8pt}
\caption{\label{tbl:ch3oh} Calculated CCSD(T)/haQZ vibrational frequencies for the methanol molecule in cm$^{-1}$ compared to MULTIMODE results from Bowman \emph{et al.}\cite{Bowman2007} (MM) and experimental reference data. The last row contains the mean absolute error (MAE) in cm$^{-1}$ compared to the experimental values.}
\begin{tabular}{@{}lrrrrrrrrr@{}}
\hline\hline
Mode	&	Harm.	&	Morse	&	VPT2	&	VPT2+K	&	GVPT2	&	VPT2+K+R	&	GVPT2+R	&	MM\cite{Bowman2007}	&	Exp.	\\
\hline
$\nu_1$	&	292	&	294	&	240	&	240	&	240	&	292	&	292	&	267	&	295$^a~$	\\
$\nu_2$	&	1059	&	1059	&	1033	&	1033	&	1033	&	1033	&	1014	&	1026	&	1034\cite{Rueda2005}	\\
$\nu_3$	&	1087	&	1083	&	1067	&	1067	&	1067	&	1067	&	1086	&	1080	&	1074\cite{Serrallach1974}	\\
$\nu_4$	&	1180	&	1180	&	1150	&	1150	&	1150	&	1150	&	1150	&	1156	&	1164\cite{Lees2000}	\\
$\nu_5$	&	1382	&	1381	&	1329	&	1329	&	1324	&	1329	&	1339	&	1322	&	1336\cite{Rueda2005}		\\
$\nu_6$	&	1484	&	1483	&	1449	&	1449	&	1445	&	1449	&	1446	&	1447	&	1454\cite{Serrallach1974}	\\
$\nu_7$	&	1511	&	1511	&	1469	&	1469	&	1469	&	1469	&	1469	&	1475	&	1482\cite{Serrallach1974}	\\
$\nu_8$	&	1522	&	1521	&	1475	&	1475	&	1480	&	1475	&	1478	&	1484	&	1486\cite{Serrallach1974}	\\
$\nu_9$	&	3015	&	2964	&	2848	&	2802	&	2829	&	2802	&	2829	&	2840	&	2844\cite{Serrallach1974}	\\
$\nu_{10}$	&	3075	&	3074	&	2939	&	2939	&	2914	&	2939	&	2914	&	2962	&	2967\cite{Serrallach1974}	\\
$\nu_{11}$	&	3135	&	3040	&	2990	&	2990	&	2990	&	2990	&	2990	&	2986	&	2999\cite{Serrallach1974}	\\
$\nu_{12}$	&	3864	&	3667	&	3680	&	3680	&	3680	&	3680	&	3680	&	3675	&	3684\cite{Rueda2005}		\\
\hline
$\nu_1+\nu_2$	&	1352	&	1353	&	1279	&	1279	&	1279	&	1245	&	1225	&	1235	&	1232\cite{Henningsen1983}	\\
$\nu_2+\nu_3$	&	2147	&	2143	&	2092	&	2092	&	2092	&	2092	&	2092	&	2098	&	2097\cite{Lees2002}	\\
$\nu_2+\nu_4$	&	2239	&	2239	&	2177	&	2177	&	2177	&	2177	&	2177	&	2177	&	2191\cite{Lees2002}	\\
\hline
$2\nu_2$	&	2119	&	2119	&	2058	&	2058	&	2058	&	2058	&	2058	&	2039	&	2055\cite{Rueda2005}		\\
$2\nu_3$	&	2175	&	2163	&	2128	&	2128	&	2128	&	2128	&	2128	&	2152	&	2141\cite{Serrallach1974}	\\
$2\nu_5$	&	2764	&	2760	&	2644	&	2644	&	2665	&	2644	&	2665	&	2678	&	2632\cite{Rueda2005}		\\
\hline
MAE & 69  & 50 & 14 & 16 & 17 & 11 & 13 & 11 & 0\\
\hline\hline
\end{tabular}\\
$^a$A $\rightarrow$ A in the gas phase from Ref. \citenum{Perchard2007}
\end{table*}

\section{Results and Discussion}

\subsection{Isolated Molecules}

Before discussing vibrations within dimers, we briefly illustrate the situation for calculations of isolated molecules (monomers) focusing on one of our largest monomers --- methanol. 
The obtained CCSD(T)/haQZ results of all above described flavors of approximations for vibrational frequencies are shown in Table \ref{tbl:ch3oh} for the methanol molecule. Note that we herein always enumerate the modes with increasing wave number according to the harmonic frequencies and if modes are degenerate they still retain their individual indices. This labeling scheme is also utilized in all the supporting data and we hope that in this way automatic processing of the supplied database should be fairly straightforward, especially given the quite diverse set of systems and our focus on low-frequency vibrations. The first twelve rows show the results for the fundamental vibrations, followed by three selected combination bands, and three selected overtones. As reference we utilize on one hand experimental results, on the other hand high-level calculations from Bowman \emph{et al.}\cite{Bowman2007}. These calculations are based on a computed potential energy surface at the CCSD(T)/aug-cc-pVTZ level and subsequent variational calculations within the MULTIMODE code (MM). For a consistent comparison with the MM results, we utilize the same experimental values and the same selection of overtones and combination bands as in Ref. \citenum{Bowman2007}.

Let us first discuss the performance of the harmonic approximation. On average, the harmonic approximation overestimates the vibrational frequencies by about 69 cm$^{-1}$ compared to the experimental results. Even by optimizing an empirical scaling parameter for harmonic frequencies that minimizes the MAE specifically for this system (0.967) still leads to a mean absolute error (MAE) of 27 cm$^{-1}$. In most cases, the Morse oscillators behave similar to the harmonic oscillators; among the fundamental modes only $\nu_9$ and $\nu_{11}$ show a significant improvement compared to the harmonic results, leading then to a MAE of 50 cm$^{-1}$. 

The standard VPT2 approach performs already very well with a MAE of only 14 cm$^{-1}$. In comparison, the error of the MM approach amounts to 11 cm$^{-1}$. For VPT2, the largest shown errors amount to 55 cm$^{-1}$ and 47 cm$^{-1}$ for the internal rotation $\nu_1$ and the $\nu_1+\nu_2$ combination band, respectively. 
CFOUR has identified a resonance of $\nu_9$ with $2\nu_6$ and $2\nu_8$, respectively, which is then accounted for in VPT2+K---leading here to a MAE of 16 cm$^{-1}$. While this resonance treatment worsens in this case the frequency for $\nu_9$, the frequency of the two overtones (not listed in the table) improves by about 21 and 15 cm$^{-1}$, respectively.
GVPT2 has taken 8 resonances into account, leading for some modes to changes up to 27 cm$^{-1}$, while many other frequencies are virtually identical to VPT2. This results then in a MAE of 17  cm$^{-1}$. 

Among the fundamental modes, the lowest-frequency mode $\nu_1$ is in all three so far discussed VPT2 variants underestimated by about 50 cm$^{-1}$. This mode describes the internal rotation of the hydroxyl group against the methyl group and hence already constitutes a somewhat problematic large-amplitude motion. Therefore, we have decoupled this mode and replaced it with a hindered rotor, leading then to the VPT2+K+R and GVPT2+R results.This results in slightly improved MAE values of 11 and 13 cm$^{-1}$, respectively. It can be seen that the MAE of our best result is in this case identical to that of the MM approach, leading to a very good description of the rather small but already quite complex methanol molecule. We note here that the MM approach would be superior by 2 cm$^{-1}$ when the lowest-frequency mode is excluded.

\subsection{Hydrogen-Bonded Dimers of Polar Molecules}

After having established what to expect in terms of accuracy for a single molecule, we now move on to the discussion of selected dimers within the proposed V30 data set. Therefore, we have selected three representatives for each of the three categories, focusing on systems for which experimental data is available for all or most fundamental modes. In the remainder we will focus on fundamental vibrations, but all calculated overtones and combinations are available in the database within the Supporting Data\cite{SI}. For the class of hydrogen-bonded dimers we have selected the two systems, which were already studied with VPT2 at the CCSD(T)/haQZ level by Howard \emph{et al.}\cite{Howard2014} --- \hfdim and H$_2$O$\ldots$H$_2$O plus additionally HF$\ldots$HCN. \\

\subsubsection{\waterdim}

\begin{table*}
\setlength{\tabcolsep}{12pt}
\caption{\label{tbl:h2oh2o} Fundamental frequencies of \waterdim calculated with CCSD(T)/haQZ in cm$^{-1}$ compared to the corresponding results from Howard et al.\cite{Howard2014} and experimental values. Since no resonance treatment in terms of VPT2+K was performed  and no mode replaced by a hindered rotor, the respective +K and +R methods yield identical results.}
\begin{tabular}{@{}lrrrrrrrr@{}}
\hline\hline
Mode	&	Harm.	&	Morse	&	Morse(0.25)$^a$	& Morse(0.5)$^a$ &	VPT2	&	GVPT2	&	Calc. Ref. \citenum{Howard2014}	&	Exp.	\\
\hline
$\nu_1$ 	&	126	&	129	&	194	&	316 & 84	&	84	&	79	&	88\cite{Braly2000}	\\
$\nu_2$ 	&	144	&	146	&	193	&	293 & 118	&	117	&	111	&	108\cite{Braly2000}	\\
$\nu_3$ 	&	149	&	149	&	177	&	241 & 110	&	110	&	103	&	103\cite{Braly2000}	\\
$\nu_4$ 	&	185	&	181	&	187	&   201 & 146	&	146	&	143	&	143\cite{Keutsch2003}	\\
$\nu_5$ 	&	352	&	350	&	359	&	386 & 299	&	293	&	293	&	311\cite{Bouteiller2004}	\\
$\nu_6$ 	&	615	&	616	&	622	&	642 & 495	&	495	&	495	&	523\cite{Bouteiller2004}	\\
$\nu_7$ 	&	1651	&	1647	&	1647	&	1646 & 1605	&	1601	&	1599	&	1599\cite{Bouteiller2004}	\\
$\nu_8$ 	&	1671	&	1667	&	1667	&	1666 & 1617	&	1620	&	1616	&	1616\cite{Bouteiller2004}	\\
$\nu_9$ 	&	3750	&	3601	&	3601	&	3602 & 3604	&	3592	&	3605	&	3602\cite{Otto2014}	\\
$\nu_{10}$ 	&	3826	&	3734	&	3734	&	3735 & 3650	&	3659	&	3650	&	3651\cite{Otto2014}	\\
$\nu_{11}$ 	&	3913	&	3795	&	3795	&	3797 & 3727	&	3731	&	3727	&	3730\cite{Otto2014}	\\
$\nu_{12}$ 	&	3932	&	3932	&	3934	&	3940 & 3742	&	3743	&	3747	&	3745\cite{Otto2014}	\\
\hline
MAE & 91 & 61 & 74 & 104 & 7 & 8 & 6 & 0\\
\hline\hline
\end{tabular}\\
$^a$ Morse fits obtained by using larger displacements corresponding to an amplitude of up to 0.25 and 0.5 in reduced normal coordinates, respectively.\\
\end{table*}
\begin{figure}
\includegraphics[width=0.7\columnwidth]{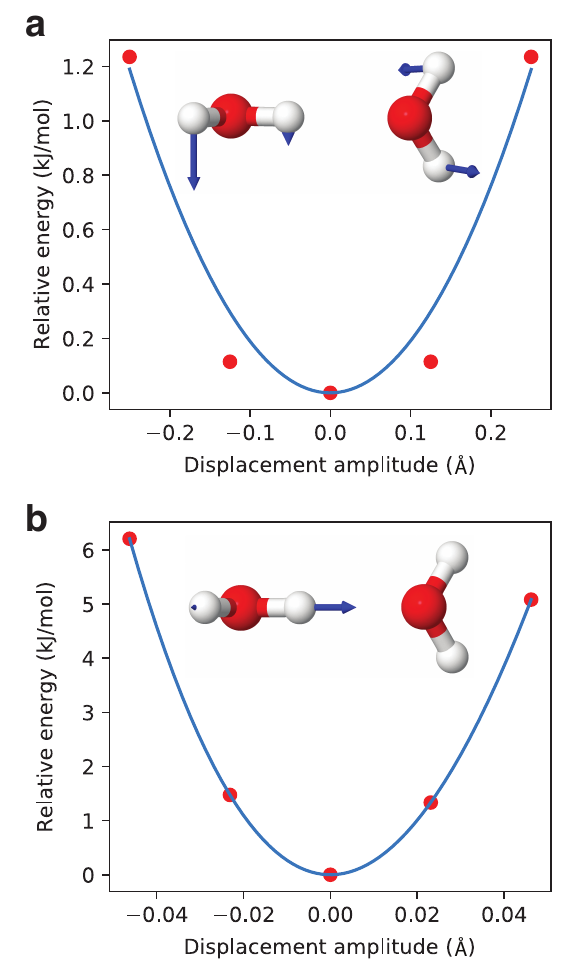}%
\caption{\label{fig:h2o} Visualization of Morse fits for $\nu_1$ (a) and $\nu_9$ (b) of the water dimer H$_2$O$\ldots$H$_2$O. 
The calculated energies at displaced geometries are shown as red dots and the Morse fit as blue curve.
Note that these fits correspond to the larger displacements in column Morse(0.5) in Table \ref{tbl:h2oh2o}. 
The corresponding normal modes are also visualized (blue arrows).}%
\end{figure}

First, we discuss the water dimer (H$_2$O$\ldots$H$_2$O), since it is arguable the most studied hydrogen-bonded dimer.
Our obtained results are listed in Table \ref{tbl:h2oh2o}.
In contrast to the above-mentioned methanol example, we observe now not only internal modes of one molecule, but also intermolecular modes ($\nu_1$ to $\nu_6$), which consist of intermolecular translations and rotations.
One example of such a mode is visualized in Fig. \ref{fig:h2o}(a).
Like for internal modes, the harmonic approximation also typically overestimates the vibrational frequencies of the intermolecular modes, which leads to a MAE of 91 cm$^{-1}$ compared to experimental reference data; the largest observed error amounts to 187 cm$^{-1}$ for $\nu_{12}$. By utilizing for the harmonic calculations an empirical scaling parameter specifically tuned to this system (0.954), the maximal accuracy that can be obtained amounts to a MAE of 25 cm$^{-1}$. Note that a thorough study regarding DFT VPT2 results and scaling of frequencies was performed by Buczek et al.\cite{Buczek2021}, which also included the water dimer.

Utilizing our Morse oscillator approach leads to an improvement over the standard harmonic approximation with a MAE of 61 cm$^{-1}$. It can be seen that in this case, the results for the intermolecular modes are basically identical to the harmonic ones, significant differences occur here only for modes $\nu_9$, $\nu_{10}$, $\nu_{11}$. These three modes describe bond-stretching vibrations under participation of the hydrogen-bond-donor water molecule. In $\nu_9$, [see Fig. \ref{fig:h2o}(b)] the dominant motion is the O---H bond stretch within the hydrogen bridge. For this vibrational mode, the Morse approach virtually exactly reproduces the experimental results, and the corresponding Morse fit is visualized in Fig. \ref{fig:h2o}(b).

Since we are utilizing quite small displacements for our standard Morse fits, we have also studied the effect of larger displacements on the vibrational frequencies of H$_2$O$\ldots$H$_2$O. As can be seen in Table \ref{tbl:h2oh2o}, increasing the maximal displacement to correspond to an amplitude of 0.25 in reduced normal coordinates increases the MAE to 74  cm$^{-1}$ and increasing it further to 0.5 makes the description with a MAE of 104 cm$^{-1}$ even worse than the harmonic approximation. Increasing the displacement for the Morse fits leads to significant differences for the low-frequency intermolecular modes, while the higher-frequency modes remain quite similar. 

In order to illustrate this problem for the intermolecular modes, the Morse fit of $\nu_1$ utilizing a large displacement (0.5) is shown in Fig. \ref{fig:h2o}(a). It can be seen that here the Morse potential resembles actually a harmonic potential, but two of the five points are located quite far away from the fit. When performing this fit for our standard displacement, all five points are located on top of an essentially harmonic potential. This problem arises since we are utilizing cartesian normal coordinates and the intermolecular modes involve so-called large-amplitude motions, which cannot be correctly described in this coordinate system. While cartesian modes are quite accurate for very small displacements, following them here will not result in a proper rotation of the water molecules and hence results in our Morse case in a significant overestimation of the corresponding vibrational frequencies. This problem can be avoided by utilizing proper curvilinear internal coordinates, which is however not trivial in an automatic fashion for intermolecular modes. 

In general, cartesian normal coordinates are problematic for large-amplitude motions for all methods relying on the harmonic approximation as a starting point. For our Morse fits, we try to minimize this problem by using in the standard approach only very small displacements. Monteiro et al.\cite{Monteiro2016} have for instance shown with MP2/aug-cc-pVTZ calculations for \waterdim that the large error of a VSCF approach based on cartesian normal modes can be significantly reduced by utilizing internal coordinates.

Now, let us move on to the discussion of the VPT2 results. For \waterdim , CFOUR has not identified any resonances and the resulting standard VPT2 fundamental frequencies are already very accurate in this case, with a MAE of only 7 wave numbers compared to experimental results. It can be seen in Table \ref{tbl:h2oh2o} that also the low-frequency modes are captured quite accurately. Although we are also utilizing cartesian normal modes as the unperturbed system, this seems not to be an issue for the water dimer, probably because of the quite strong interaction between the two water molecules due to the hydrogen bond. However, we note here that large-amplitude motions can also be problematic for VPT2, as will be discussed for other systems later on.

As mentioned above, Howard et al.\cite{Howard2014} have already studied the water dimer with VPT2 employing also CCSD(T)/haQZ. Hence, we compare these results in Table. \ref{tbl:h2oh2o} and we also use for consistency the same experimental reference data. In terms of the MAE w.r.t. experiment, the results of Howard et al. are with 6 cm$^{-1}$ about one wave number closer to experiment than our VPT2 results. However, we observe small frequency differences between these two VPT2 calculations.
To explain this small discrepancy, we have tested for haTZ calculations different displacement step sizes and compared the results with haTZ results from Ref. \citenum{Howard2014} (see Table S2 in the Supporting Information). With the default step size in CFOUR (0.05) we observe differences up to 10 cm$^{-1}$, with a MAE of about 3 cm$^{-1}$. Here, the main differences originate from the lower-frequency modes, while the higher-frequency modes agree within 1 cm$^{-1}$. We have now tested step sizes ranging from 0.01 to 0.1.  While the higher-frequency modes remain basically unaffected by the change in step size, the three lowest-frequency modes change significantly with the interval being 98 - 57 cm$^{-1}$ for $\nu_1$, 125 - 100 cm$^{-1}$ for $\nu_2$, and 115 - 102 cm$^{-1}$ for $\nu_3$, respectively.  It can be seen that by modification of the utilized step size to 0.08, it is possible to reproduce the haTZ results from Ref. \citenum{Howard2014} within a MAE of 1 cm$^{-1}$. 

Finally, we have also computed the fundamental frequencies with GVPT2, which leads to a MAE of 8 cm$^{-1}$. This approach has identified one Fermi resonance and several 1-1 and 2-2 Darling-Dennison resonances; we observe some shifts for higher frequencies with the largest difference to VPT2 being 12 cm$^{-1}$ for $\nu_9$. All discussed second-order perturbation variants yield excellent results for H$_2$O$\ldots$H$_2$O, with a maximal MAE of 8 cm$^{-1}$ compared to experimental data. For comparison, ideally re-scaled harmonic frequencies would lead to a more then threefold larger error.\\

\subsubsection{\hfdim}

The second example for hydrogen-bonded dimers is \hfdim and the results are listed in Table \ref{tbl:hfhf}. 
It can be seen that the harmonic approximation shows again a significant overestimation of the fundamental frequencies with a MAE of 100 cm$^{-1}$, which could by ideal empirical scaling (0.958) be reduced to 38 cm$^{-1}$. Utilizing Morse oscillators reduces the error of the standard harmonic approximation to 52 cm$^{-1}$. Again, the Morse oscillators mainly improve the description of bond stretches ($\nu_5$, $\nu_6$). 

The VPT2 description of the fundamental modes is also here already very accurate, with a MAE of 9 cm$^{-1}$. CFOUR has not identified any resonances, therefore no treatment for VPT2+K is necessary. Due to the quite strong interaction between the two HF molecules, we observe here also no issue with large-amplitude motions and no hindered rotor treatment is necessary. Since \hfdim was also studied by Howard et al.\cite{Howard2014}, we again compare to their VPT2 results, which have a MAE of 8 cm$^{-1}$ w.r.t. the experimental data. As for the \waterdim we observe some differences between the calculated frequencies resulting on average to about 3 cm$^{-1}$. In Table S3, we compare haTZ calculations utilizing different displacements with the haTZ results of Howard et al. As for the water dimer, varying the displacements enables a reproduction of the results with the MAE at a displacement of 0.04 being less than 1 cm$^{-1}$. In the GVPT2 case, there exists one 1-1 Darling-Dennison resonance, which leads only to a very minor change in fundamental frequencies up to 0.5 cm$^{-1}$ and hence results like VPT2 in a MAE of 9 cm$^{-1}$.

\begin{table}
\setlength{\tabcolsep}{3pt}
\caption{\label{tbl:hfhf} Fundamental frequencies of \hfdim calculated with CCSD(T)/haQZ in cm$^{-1}$ compared to the corresponding results from Howard et al.\cite{Howard2014} and experimental values.}
\begin{tabular}{@{}lrrrrrr@{}}
\hline\hline
Mode    & Harm. & Morse & VPT2 & GVPT2 & Calc. Ref. \citenum{Howard2014} & Exp. \\
\hline
$\nu_1$ 	&	161	&	156	&	131	&	131	&	132	&	125\cite{Quack1991}	\\
$\nu_2$ 	&	218	&	219	&	170	&	171	&	172	&	161\cite{Quack1990}	\\
$\nu_3$ 	&	466	&	467	&	399	&	399	&	389	&	380\cite{Quack1996}	\\
$\nu_4$ 	&	568	&	552	&	458	&	458	&	453	&	475\cite{Anderson1996}	\\
$\nu_5$ 	&	4025	&	3822	&	3871	&	3871	&	3869	&	3868\cite{Pine1984}	\\
$\nu_6$ 	&	4104	&	3918	&	3929	&	3930	&	3930	&	3931\cite{Pine1984}	\\
\hline
MAE & 100 & 52 & 9 & 9 & 8 & 0\\
\hline\hline
\end{tabular}
\end{table}

\subsubsection{HF$\ldots$HCN}

As a third example for hydrogen-bonded dimers we discuss HF$\ldots$HCN (see Table \ref{hfhcn}). Note that we combine in the main text tables (nearly) degenerate modes for the comparison with experimental data and show here the average value for those modes (full data is available in the Supporting Information). Also, these modes count only one time for the MAE calculation.  
We observe the same trends as before, with the harmonic approximation significantly overestimating the fundamental frequencies with a MAE of 64 cm$^{-1}$ while utilizing independent Morse oscillators brings the MAE down to 29 cm$^{-1}$. A hypothetical ideal empiric scaling (0.964) would reduce the MAE of the harmonic approximation to 24 cm$^{-1}$. In this case VPT2 and GVPT2 yield excellent results with the MAE being only 5 cm$^{-1}$ and the maximal error 10 cm$^{-1}$ in both cases. GVPT2 yields for the fundamental modes here virtually identical results to VPT2 since only two 2-2 Darling-Dennison resonances were identified. Given the quite strong hydrogen bond, also here no hindered rotor treatment is necessary.

\begin{table}
\setlength{\tabcolsep}{8pt}
\caption{\label{hfhcn} Fundamental frequencies of HF$\ldots$HCN calculated with CCSD(T)/haQZ in cm$^{-1}$ compared to  experimental values.}
\begin{tabular}{@{}lrrrrr@{}}
\hline\hline
Mode &	Harm.	&	Morse  &  VPT2 & GVPT2	& Exp.	\\
\hline
$\nu_{1,2}$ &   84 &    84 &   77  &   76 & 74\cite{Dayton1988} \\
$\nu_3$     &  187 &   178 &   173 &  173 & 168\cite{Bender1987}\\
$\nu_{4,5}$ &  630 &   630 &   560 &  560 & 550\cite{Wofford1986}\\
$\nu_{6,7}$ &  737 &   737 &   729 &  728 & 727\cite{Wofford1986b}\\
$\nu_8$     & 2152 &  2132 &  2117 & 2117 & 2121\cite{Wofford1985}\\
$\nu_9$     & 3434 &  3325 &  3303 & 3303 & 3310\cite{Kyr1986} \\
$\nu_{10}$  & 3887 &  3648 &  3722 & 3722 & 3716\cite{Kyr1983}\\
\hline
MAE & 64 & 29 & 5 & 5 &  0\\

\hline\hline
\end{tabular}\\
\end{table}

\subsection{Complexes of Polar and Non-Polar Molecules}

\subsubsection{HF$\ldots$N$_2$}

Now we turn to the discussion of complexes between a polar molecule and a non-polar molecule, with our first example being HF$\ldots$N$_2$ (see Table \ref{tab:HF_N2}). This example looks qualitatively just like the systems involving two polar molecules with the harmonic approximation, an ideally scaled harmonic approximation (0.961), and the Morse oscillators  yielding MAEs of 83 cm$^{-1}$, 37 cm$^{-1}$, and 55 cm$^{-1}$, respectively. Both VPT2 variants yield virtually identical results with the MAE being again only 5 cm$^{-1}$ in both cases. The largest deviation from experiment is found for the H--F bend with 13 cm$^{-1}$, while the other modes are described within 4 cm$^{-1}$. Hence, no hindered rotor treatment is necessary.

\begin{table}[h]
\centering
\caption{Fundamental frequencies of HF\ldots N$_2$ calculated with CCSD(T)/haQZ in cm$^{-1}$ compared to experimental values. Only fundamental modes with available reference values are listed.}
\label{tab:HF_N2}
\begin{tabular}{@{}lrrrrr@{}}
\hline\hline
Mode & Harm. & Morse & VPT2 & GVPT2 & Exp. \\
\hline
$\nu_{1,2}$ & 77 & 78 & 61 & 62 & 59\cite{Tsang1996}\\
$\nu_{3}$ & 118 & 109 & 89 & 90  & 91\cite{Lovejoy1987}\\
$\nu_{4,5}$ & 384 & 386 & 272 & 272 & 259\cite{Tsang1996}\\
$\nu_{7}$ & 4079 & 3863 & 3922 & 3922 & 3918\cite{Lovejoy1987}\\
\hline
MAE & 83 & 55 & 5 & 5 & 0\\
\hline\hline
\end{tabular}
\end{table}

\subsubsection{\hfar}

The second example in this category is \hfar (see Table \ref{hfar}). Like in the examples above, the harmonic approximation leads to a significant overestimation of vibrational frequencies with a MAE of 94 cm$^{-1}$ and ideal empirical scaling (0.958) would bring that error down to 34 cm$^{-1}$. The Morse oscillator approach has a MAE of 50 cm$^{-1}$. Here, it improves the description of $\nu_1$ and $\nu_4$, while further overestimating $\nu_{2,3}$.

\begin{table}
\setlength{\tabcolsep}{1pt}
\caption{\label{hfar} Fundamental frequencies of \hfar calculated with CCSD(T)/haQZ in cm$^{-1}$ compared to  experimental values.}
\begin{tabular}{@{}lrrrrrrr@{}}
\hline\hline
Mode	&	Harm. 	&	Morse	&	VPT2	&	GVPT2	&	VPT2+R	&	GVPT2+R	&	Exp.	\\
\hline
$\nu_1$ 	&	62	&	52	&	41	&	41	&	41	&	41	&	39\cite{Lovejoy1989}	\\
$\nu_{2,3}$ 	&	158	&	165	&	-12	&	-10	&	77	&	77	&	70\cite{Lovejoy1986}	\\
$\nu_4$ 	&	4124	&	3912	&	3954	&	3956	&	3954	&	3956	&	3952\cite{Lovejoy1986}	\\
\hline
MAE & 94 & 50 & 28 & 29 & 3 & 4 & 0\\
\hline\hline
\end{tabular}
\end{table}

\begin{figure*}[!]
\includegraphics[width=\textwidth]{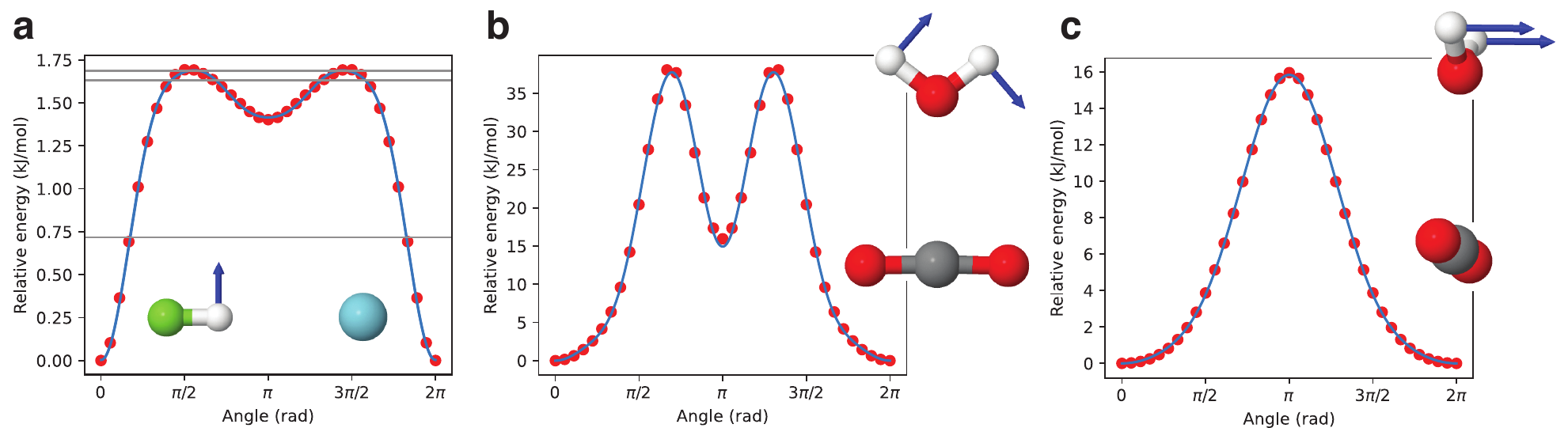}%
\caption{\label{fig:rotor1}Visualization of modes representing large-amplitude motions and relative CCSD(T)/haQZ energies for the corresponding hindered rotations for HF$\ldots$Ar (a) and \watercoo (b, c). The gray lines in a) represent the obtained hindered rotor eigenvalues.}%
\end{figure*}

While VPT2 very accurately describes here the lowest frequency mode as well as the highest frequency mode, we observe for $\nu_{2,3}$ an imaginary frequency of -12 cm$^{-1}$. This leads then to a MAE of 28 cm$^{-1}$. CFOUR has not identified any resonances, so VPT2+K yields the same result. In the GVPT2 case we have one 2-2 Darling-Dennison resonance leading virtually to the same results with a MAE of 29 cm$^{-1}$ and an imaginary frequency. 

The problematic mode $\nu_{2,3}$ describes basically a hindered rotation of the hydrogen atom around the fluorine atom, or more precisely around the HF center-of-mass (see Fig. \ref{fig:rotor1}(a)). This is a large-amplitude motion which cannot be correctly captured by cartesian normal coordinates. We have tested also different displacements for VPT2 with haTZ. Utilizing smaller displacements (0.01) results in a frequency of about 10 cm$^{-1}$, which is still far off the experimental result, while increasing the displacement to 0.1 leads to about -100 cm$^{-1}$. 

Therefore, we decouple $\nu_{2,3}$ from the other modes and treat this degenerate mode as an one-dimensional hindered rotor. The calculated potential for this hindered rotation is plotted in see Fig. \ref{fig:rotor1}(a). It can be seen that the rotational barrier is very small with 1.7 kJ/mol (142 cm$^{-1}$). Hence, we obtain the corresponding frequency from a numerical solution of the hindered rotor Hamiltonian. This leads to 77 cm$^{-1}$, which is only 7 cm$^{-1}$ off the experimental result. In contrast, obtaining such a frequency from the curvature in case of such small rotation barriers would lead to a significant overestimation (147 cm$^{-1}$). Utilizing this rotor frequency in our VPT2+R and GVPT2+R approach, reduces the MAE w.r.t. experiment to only 3 and 4 cm$^{-1}$, respectively. Here, the decoupling of $\nu_{2,3}$ did also not lead to any changes in the other two frequencies.

\subsubsection{\watercoo}

\begin{table*}
\setlength{\tabcolsep}{8.5pt}
\caption{\label{h2oco2} Fundamental frequencies of \watercoo calculated with CCSD(T)/haQZ in cm$^{-1}$ compared to the corresponding results from Wang and Bowman\cite{Wang2017} as well as experimental values from Soulard and Tremblay\cite{Soulard2015}. The first MAE value corresponds to comparing with all experimental frequencies and the second value compares only frequencies with available MULTIMODE results. Only fundamental modes with available reference values are listed.}
\begin{tabular}{@{}lrrrrrrrrr@{}}
\hline\hline
Mode	&	Harm.	&	Morse	&	VPT2	&	VPT2+K	&	GVPT2	&	VPT2+K+R	&	GVPT2+R	&	MULTIMODE\cite{Wang2017}	&		Exp.\cite{Soulard2015}	\\
\hline
$\nu_1$ 	&	24	&	45	&	-84	&	-84	&	-99	&	152	&	152	&		&		167	\\
$\nu_2$ 	&	87	&	100	&	43	&	43	&	36	&	103	&	103	&		&		103	\\
$\nu_6$ 	&	660	&	660	&	657	&	657	&	658	&	657	&	658	&	660	&		659	\\
$\nu_7$ 	&	674	&	674	&	670	&	670	&	669	&	670	&	669	&	676	&		672	\\
$\nu_8$ 	&	1350	&	1344	&	1919	&	1276	&	1259	&	1276	&	1259	&	1285	&			\\
$\nu_9$ 	&	1646	&	1641	&	1598	&	1598	&	1596	&	1598	&	1596	&	1583	&		1595	\\
$\nu_{10}$ 	&	2389	&	2386	&	2342	&	2342	&	2343	&	2342	&	2343	&	2348	&		2348	\\
$\nu_{11}$ 	&	3831	&	3735	&	3649	&	3649	&	3652	&	3649	&	3652	&	3663	&		3665	\\
$\nu_{12}$ 	&	3941	&	3941	&	3746	&	3746	&	3749	&	3746	&	3749	&	3759	&		3757	\\

\hline
MAE & 76/74 & 58/57 & 44/7 & 44/7 & 46/5 & 7/7 & 6/5 & -/3 & 0/0\\
\hline\hline
\end{tabular}\\
\end{table*}

Next, we discuss the results of \watercoo, which are listed in Table \ref{h2oco2} for all modes with available reference data. In addition to experimental results\cite{Soulard2015}, we can also compare to high-level MULTIMODE calculations from Wang and Bowman\cite{Wang2017}. First, we compare all our calculated results with the available experimental data. 
The harmonic approximation leads to a large MAE of 76 cm$^{-1}$, which could be reduced by an ideal empirical scaling (0.957) to 39 cm$^{-1}$. Utilizing independent Morse oscillators again improves upon the standard harmonic approximation with a MAE of 58 cm$^{-1}$ and in this case improved also the two lowest-frequency modes. 
The standard VPT2 results initially have a MAE of 44 cm$^{-1}$, which is due to the fact that the two lowest-frequency modes are badly described due to the involved large amplitude motions. CFOUR identified here resonances between $2\nu_6$, $2\nu_7$, and $\nu_8$, which then leads to the VPT2+K results showing a significant modification of $\nu_8$. GVPT2 yields slightly different results due to more resonances but shows the same problem with the two lowest-frequency modes with a MAE of 46 cm$^{-1}$.

Both problematic modes and the corresponding fit for the hindered rotation are shown in Fig. \ref{fig:rotor1}. It can be seen that in both cases the water molecule is rotating around its center of mass and the rotational barriers are fairly large with at least 16 kJ/mol. Therefore, we obtain the associated frequency from the curvature of the fitted potential, which leads for $\nu_1$ and $\nu_2$ to 152 cm$^{-1}$ and 103 cm$^{-1}$, respectively. This results now in a significant improvement of the result with a MAE of only 7 cm$^{-1}$ for VPT2+K+R and 6 cm$^{-1}$ for GVPT2+R. 

Frequencies from high-quality reference calculations using the MULTIMODE approach are only available for higher-frequency vibrations and have a MAE of 3 cm$^{-1}$ compared to experiment. In our case, we can reach with VPT2+K+R and GVPT2+R a MAE of 7 and 5 cm$^{-1}$, respectively, when only taking into account frequencies also available in the MULTIMODE approach. This illustrates that the utilized VPT2 methodologies can lead to extremely accurate results with errors only slightly larger than the high-quality MULTIMODE approach.

\subsection{vdW-Bound Dimers of Non-Polar Molecules}

\subsubsection{\methanear}

Next, we turn to the discussion of dimers consisting of only non-polar molecules. Our first representative example is \methanear and the obtained results are listed in Table \ref{ch4ar}.  While experimental results are available for higher-frequency modes, we have to rely on high-level theoretical estimates\cite{Ferenc2018} for $\nu_1$ and $\nu_{2,3}$. 

\begin{table*}
\setlength{\tabcolsep}{8pt}
\caption{\label{ch4ar} Fundamental frequencies of \methanear calculated with CCSD(T)/haQZ in cm$^{-1}$ compared to reference values. Only fundamental modes with available reference values are listed.}
\begin{tabular}{@{}lrrrrrrr@{}}
\hline\hline
Mode &	Harm.	&	Morse	&	VPT2	&	GVPT2	&	VPT2+R	&	GVPT2+R	&	Reference	\\
\hline
$\nu_1$ 	&	47	&	37	&	22	&	21	&	22	&	21	&	29\cite{Ferenc2018}	\\
$\nu_{2,3}$ 	&	61	&	65	&	-5	&	-5	&	40	&	40	&	32\cite{Ferenc2018}	\\
$\nu_{4,5,6}$ 	&	1345	&	1344	&	1308	&	1308	&	1308	&	1308	&	1311\cite{Pak1998}	\\
$\nu_{10,11,12}$ 	&	3153	&	3129	&	3010	&	3010	&	3010	&	3010	&	3016\cite{Chamberland1970}	\\

\hline
MAE & 55 & 47 & 13 & 13 & 6 & 6 & 0\\
\hline\hline
\end{tabular}
\end{table*}

\begin{figure}
\includegraphics[width=\columnwidth]{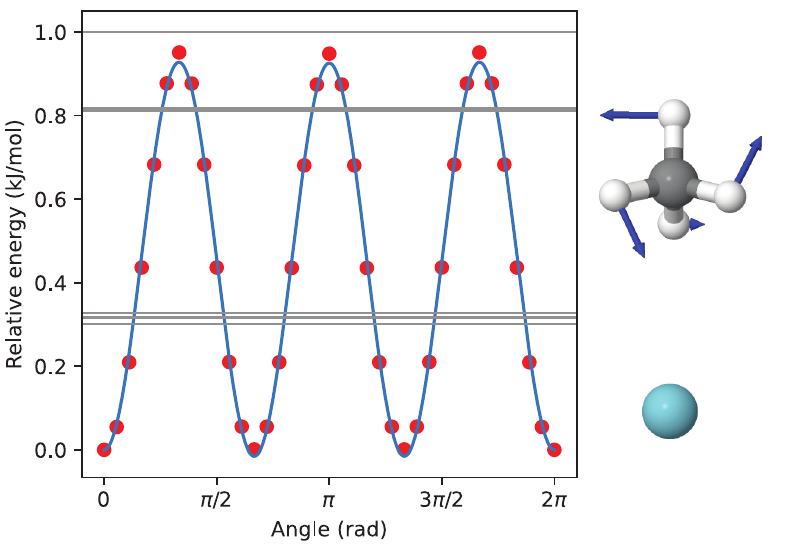}%
\caption{\label{fig:rotor2}Visualization of a CH$_4\ldots$Ar mode representing a large-amplitude motion and relative CCSD(T)/haQZ energies for the corresponding hindered rotation. The gray lines represent the obtained hindered rotor eigenvalues.}%
\end{figure}

The harmonic approximation yields here an error of 55 cm$^{-1}$ and an ideal empirical scaling (0.957) would reduce it to 17 cm$^{-1}$. The Morse oscillators yield quite similar results as the harmonic approximation with a MAE of 47  cm$^{-1}$. Standard VPT2 and GVPT2 have each a MAE of 13 cm$^{-1}$ and  $\nu_{2,3}$ have imaginary frequencies of -5 cm$^{-1}$. These modes describe hindered rotations of the methane molecule around its center of mass (see Fig. \ref{fig:rotor2}). It can be seen that the rotation barrier is extremely small and amounts to less than 1 kJ/mol or 84 cm$^{-1}$. Therefore, we utilize the numerical solution of the hindered rotor Hamiltonian, which yields a frequency of 40 cm$^{-1}$, which agrees within 8 cm$^{-1}$ with the calculated reference value. In contrast, deriving this frequency from the curvature of the fitted potential would yield 58 cm$^{-1}$. The resulting VPT2+R and GVPT2+R approaches have each then a MAE of 6 cm$^{-1}$ compared to the utilized reference values.

\subsubsection{CO$_2\ldots$Ar}

While in many cases within this category a hindered rotor treatment is necessary for some modes, we also find systems like CO$_2\ldots$Ar and CO$_2\ldots$CO$_2$, where the intermolecular modes can already be well described by the standard methods. Table \ref{tab:CO2_Ar} shows the resulting fundamental modes of CO$_2\ldots$Ar. In this case, the harmonic approximation already provides good results with a MAE of 11 cm$^{-1}$, which could theoretically be improved to 6 cm$^{-1}$ by ideal empirical scaling (0.984). Utilizing Morse oscillators instead of harmonic oscillators yields virtually no difference for CO$_2\ldots$Ar. By improving the methodology to  VPT2(+K) or GVPT2, the MAE can be further reduced to 3 and 5 cm$^{-1}$, respectively. In the VPT2+K case resonances between $\nu_5$, $2\nu_3$, and $2\nu_4$ are considered, yielding to a significant shift of $\nu_5$ compared to standard VPT2. However, since there is no experimental observation of this mode, this does not affect the shown MAE. 

\begin{table}
\centering
\caption{Fundamental frequencies of CO$_2$\ldots Ar calculated with CCSD(T)/haQZ in cm$^{-1}$ compared to experimental values.}
\label{tab:CO2_Ar}
\begin{tabular}{@{}lrrrrrr@{}}
\hline\hline
Mode & Harm. & Morse & VPT2 & VPT2+K & GVPT2 & Exp.\\
\hline
$\nu_{1}$ & 35 & 35 & 28 & 28 & 29 & 28\cite{Sharpe1991}\\
$\nu_{2}$ & 44 & 44 & 35 & 35 & 27 &38\cite{Steed1979}\\
$\nu_{3}$ & 668 & 668 & 664 & 664 & 664 & 667\cite{Gartner2020}\\
$\nu_{4}$ & 669 & 669 & 666 & 666 & 666 & 668\cite{Gartner2020}\\
$\nu_{5}$ & 1349 & 1343 & 1575 & 1386 & 1371 &\\
$\nu_{6}$ & 2387 & 2387 & 2340 & 2340 & 2341 & 2349\cite{Randall1988}\\
\hline
MAE & 11 & 11 & 3 & 3 & 5& 0\\

\hline\hline
\end{tabular}\\
\end{table}

\subsubsection{\coodim}

Finally, we discuss the CO$_2\ldots$CO$_2$ dimer, for which the low-frequency modes are already quite well described with the standard methods. Table \ref{coo} lists the first four intermolecular modes, for which experimental data is available. As for CO$_2$\ldots Ar, the harmonic approximation already yields quite accurate results with a MAE of 10 cm$^{-1}$ and Morse oscillators reduce the error to 8 cm$^{-1}$. For the vibrational perturbation methods we utilize here the harmonic frequencies obtained with CCSD(T)/haQZ while the cubic and quartic force constants have been obtained with the smaller haTZ basis set. The standard VPT2 approach yields here excellent results with a MAE of only 3 cm$^{-1}$. In the GVPT2 case the quite extensive resonance treatment significantly reduces the frequencies of modes $\nu_1$ and $\nu_3$, leading then to a MAE of 7 cm$^{-1}$.

\begin{table}
\setlength{\tabcolsep}{8pt}
\caption{\label{coo} Low-frequency fundamental frequencies of \coodim calculated with CCSD(T)/haQZ in cm$^{-1}$ compared to  experimental values\cite{NoroozOliaee2016}. For VPT2/GVPT2 haQZ was used for the harmonic frequencies and anharmonic force constants were calculated with haTZ.}
\begin{tabular}{@{}lrrrrr@{}}
\hline\hline
Mode &	Harm.	&	Morse	&	VPT2	&	GVPT2	& Exp.\cite{NoroozOliaee2016}	\\
\hline
$\nu_1$ &	 26 & 26 & 26 & 19 & 23\\
$\nu_2$ &	 29 & 29 & 24 & 24 & 22\\
$\nu_3$ &	 47 & 44 & 34 & 17 & 32\\
$\nu_4$ &	108 & 101 & 95 & 100 & 92\\
\hline
MAE & 10 & 8 & 3 & 7  & 0 \\
\hline\hline
\end{tabular}\\
\end{table}

\subsection{Frequency Shifts}

So far we have only discussed the agreement of absolute vibrational frequencies. However, often one is not only interested in the absolute values but rather frequency shifts of individual vibrations in dimers or generally upon solvation. A recent in-dept study\cite{Fischer2023} --- the so-called HyDRA challenge --- focused for instance on the calculation of water donor stretching vibrations in monohydrates of organic molecules.
In our V30 set the most abundant intermolecular vibration is the HF stretch. Therefore, we compare as example in Table \ref{tab:shift} our calculated intermolecular HF stretch frequency shifts compared to the isolated HF molecule for 8 out of 9 HF-containing dimers, for which experimental reference data could be found. In all cases, the HF stretching frequency in the dimer is shifted to lower wave numbers compared to the monomer. In this case, the harmonic approximation either overestimates or underestimates the shift, leading to a MAE of 19 cm$^{-1}$. The Morse oscillators can this time not improve upon the harmonic approximation with a MAE of 23 cm$^{-1}$. Utilizing vibrational perturbation theory (VPT2 or GVPT2) results in a significant improvement in the shift description, with MAEs of only 4-6 cm$^{-1}$. Note that the rather large difference between VPT2+K and GVPT2 for HF$\ldots$NH$_3$ is due to a Darling-Dennison resonance within the GVPT2 dimer calculation.

\begin{table}
\centering
\caption{Frequency shifts in cm$^{-1}$ of the (intermolecular) HF stretching vibration in several dimers compared to the isolated molecule. For the experimental results 3961 cm$^{-1}$ (Ref. \citenum{LeRoy1999}) was utilized for the frequency of the isolated HF molecule.}
\label{tab:shift}
\begin{tabular}{@{}lrrrrr@{}}
\hline\hline
System & Harm. & Morse & VPT2 & GVPT2 & Exp.\\
\hline
HF$\ldots$NH$_3$	&	-708	&	-778	&	-725	&	-761	&	-746\cite{Johnson1982}	\\
HF$\ldots$H$_2$O	&	-362	&	-364	&	-327	&	-340	&	-327\cite{Bulychev2005}	\\
HF$\ldots$HCN	&	-255	&	-283	&	-241	&	-241	&	-245\cite{Kyr1983}	\\
HF$\ldots$HF	&	-117	&	-109	&	-92	&	-92	&	-93\cite{Pine1984}	\\
HF$\ldots$C$_2$H$_2$	&	-175	&	-194	&	-166	&	-173	&	-167\cite{Douberly2003}	\\
HF$\ldots$N$_2$	&	-63	&	-68	&	-41	&	-41	&	-43\cite{Lovejoy1987}\\
HF$\ldots$CH$_4$	&	-46	&	-49	&	-47	&	-46	&	-50\cite{Davis1987}	\\
HF$\ldots$Ar	&	-18	&	-19	&	-9	&	-7	&	-9\cite{Lovejoy1986}	\\
\hline
MAE &	19	&	23	&	4	&	6	& 0	\\
\hline\hline
\end{tabular}
\end{table}

\subsection{Overall Discussion}

The CCSD(T)/haQZ harmonic frequencies, as expected, mostly overestimate the experimentally measured fundamental frequencies of the studied dimers, but for some very low-frequency modes we also observed an underestimation. We have discussed for individual systems the respective smallest obtainable MAE by utilizing one single idealized scaling factor, which varies between 0.954 and 0.984 for all shown systems except the four low-frequency modes of CO$_2\ldots$CO$_2$; there it amounts to 0.855.  
Deriving an overall empirical scaling factor that minimized the MAE w.r.t. experiment for all fundamental modes of dimers discussed in the main text, we obtain a MAE of 29 cm$^{-1}$ at a scaling factor of 0.958. In comparison, the canonical harmonic approximation leads to a MAE of 69 cm$^{-1}$. 

Utilization of independent Morse oscillators improved upon the canonical harmonic approximation for the shown systems. This improvement mainly resulted in a better description of some stretching vibrations and a slight improvement for several low-frequency intermolecular modes. For all systems discussed this leads to a MAE of 44 cm$^{-1}$. We note that care must be taken with the Morse approach when using displacements in cartesian normal-mode coordinates. Increasingly large displacements lead to a significant overestimation of low-frequency modes, eventually becoming worse than the harmonic approximation. This implies that the anharmonic nature of vibrational modes cannot be sufficiently captured by the simplistic scaling or Morse approach.

For vibrational perturbation theory, the accounting for resonances can be important for fundamental frequencies and the VPT2+K and GVPT2 approaches lead to a fairly similar description with MAEs of 13-14 cm$^{-1}$. While these methods produce very good results for systems with quite strong intermolecular hydrogen bonds, as also already observed by Howard et al.\cite{Howard2014}, large-amplitude motions of several low-frequency modes can cause significant problems for VPT2. Modes which describe intermolecular hindered rotations cannot be accurately captured by the utilized cartesian normal-mode coordinates. This leads often to imaginary frequencies or to a significant overestimation of the frequency. Quite a large range of values can be produced by modifying the used step size for displacements. 
Already for our polar-polar group of systems large-amplitude motions present a problem for H$_2$O$\ldots$NH$_3$, CH$_3$OH$\ldots$H$_2$O, and NH$_3\ldots$NH$_3$; for our other two system groups this issue arises for at least one mode in most systems. 

Because of this, we have decoupled these problematic low-frequency modes and described them whenever possible with a one-dimensional hindered rotor model. For the shown systems the incorporation of hindered rotors (VPT2+K+R and GVPT2+R) leads to a reduction in the MAE to 6-7 cm$^{-1}$. This error is more than four times smaller than what can be obtained by just scaling harmonic frequencies. This indicates that relatively simple rotor models can be useful for describing problematic intermolecular modes. However, we have to note that we currently identify rotors manually and rely on rigid rotations. Therefore, for larger systems, a more sophisticated and mostly automatic approach should be developed. Also, more tests are required to establish the robustness of this approach since the number of data points is still fairly limited. 

Although we have only explicitly discussed nine out of the 30 systems, all systems will serve as important benchmark data for more approximate methods such as density functional approximations. The remaining 21 systems are vital for enabling a better statistical analysis with increased chemical diversity for the present non-covalent interactions.

\section{Conclusion}

We have introduced a large reference set of benchmark vibrational frequencies --- V30 --- for molecular dimers using the gold standard of quantum chemistry --- CCSD(T) --- in combination with vibrational perturbation theory. In addition, harmonic values and results of Morse oscillators were also reported. A sufficiently large basis set of quadruple-zeta quality with diffuse functions was utilized for all harmonic and Morse calculations and for all except 4 systems also for VPT2. 
Large-amplitude motions within low-frequency modes cannot properly be described by canonical VPT2 approaches, which affects many of the V30 systems. Therefore, we have decoupled these problematic modes from the rest and replaced their description with a one-dimensional hindered rotor model whenever possible. 

For the nine explicitly discussed systems, 
the MAE for the harmonic approximation amounts to 69 cm$^{-1}$ and can be reduced to 29 cm$^{-1}$ when an ideal empirical scaling factor is applied. In comparison, the combination of VPT2 with one-dimensional hindered rotor models yields very accurate results with a MAE of only 6-7 cm$^{-1}$ compared to experimental data, almost competing with the accuracy of other reference data. However, further development and testing is needed to validate this approach and to extend it to larger and more complex dimers.
As computations of intermolecular interactions and their vibrational levels become evermore important, the V30 set from this work can serve as a seminal benchmark for these important frequencies as well as thermodynamic functions derived from them.

\begin{acknowledgments}
This project has received funding from the European Union’s Horizon 2020 research and innovation programme under the Marie Skłodowska-Curie grant agreement No 890300.
The computational results presented have been achieved in part using the Vienna Scientific Cluster (VSC).
\end{acknowledgments}

\end{document}